# A Respiratory Motion Analysis for Guiding Stereotactic Arrhythmia Radiotherapy Motion Management

Yuhao Wang[1], Yao Hao[2], Hongyu An[3,4], H Michael Gach[2,3,4], Clifford G. Robinson[2,5], Phillip S. Cuculich[2,5], Deshan Yang[1*]

[1]Department of Radiation Oncology, School of Medicine, Duke University
[2]Department of Radiation Oncology, School of Medicine, Washington University in Saint Louis
[3]Department of Radiology, School of Medicine, Washington University in Saint Louis
[4]Department of Biomedical Engineering, Washington University in Saint Louis
[5]Department of Cardiology, School of Medicine, Washington University in Saint Louis

*Corresponding author: Deshan Yang, deshan.yang@duke.edu

## Abstract

**Background**: Stereotactic Arrhythmia Radiotherapy (STAR) treats ventricular tachycardia (VT) but requires internal target volume (ITV) expansions to compensate for cardiorespiratory motion. Respiratory 4D CTs (r4DCT) are commonly acquired for STAR patients to assess respiratory motion and determine the ITV expansion margin. Current clinical r4DCT imaging methods are limited, and the reconstructed r4DCTs suffer from unmanaged cardiac motion artifacts that affect the quantitative assessment of respiratory motion.

**Purpose**: To develop a novel image-processing method that accurately quantifies the respiratory motion of the heart in r4DCTs corrupted by cardiac motion artifacts.

**Methods**: A groupwise surface-to-surface deformable image registration (DIR) algorithm, named gCGF, was developed by combining the Coherent Point Drift (CPD) algorithm with Gaussian Mixture Models (GMM) and a Finite Element Model (FEM). A novel principal component filtering (PCF) mechanism and a spatial smoothing mechanism were developed and incorporated into gCGF to iteratively register heart contours from an average respiratory-phase CT to ten r4DCT phases while removing random cardiac motion from the cyclic respiratory motion. The performance of the groupwise DIR was quantitatively validated using 8 digital phantoms with simulated cardiac artifacts. An ablation study was conducted to compare gCGF to another comparable state-of-the-art groupwise DIR method. gCGF was applied to r4DCTs of 20 STAR patients to analyze the respiratory motion of the heart, which was computed for individual respiratory phases, with the average position of registered heart shapes as the reference position.

**Results**: Validation on digital phantoms showed that gCGF achieved a mean target registration error of 0.63±0.51 mm while successfully achieving phase smoothness and reducing cardiac motion artifacts. For each STAR patient, the heart contours of ten phases of r4DCT were registered with gCGF. Among all STAR patients, the heart's maximum and mean respiratory motion magnitudes ranged from 3.6 to 7.9 mm and 1.0 mm to 2.6 mm. The peak-to-peak motion range was from 6.2 to 14.7 mm. For VT targets, the max and mean motion magnitude ranges were 3.0 to 6.7 mm and 0.8 to 2.9 mm, respectively. The peak-to-peak range was from 4.7 to 11.8 mm. Significant dominance of the first principal component of the motion direction was observed ($p = 0$).

**Conclusions**: The gCGF surface-to-surface deformable registration algorithm was confirmed to be robust to quantify respiratory motion of the heart in the respiratory 4D CTs while mitigating cardiac motion artifacts. Respiratory motion was found to be patient-specific and predominantly in the first principal component direction. The gCGF algorithm and the results of this study can be useful to enable personalized motion management for STAR treatments and patient-specific optimization of the ITV margins.

# 1 Introduction

Stereotactic Arrhythmia Radiotherapy (STAR)[1] has been developed to treat drug-refractory and recurrent ventricular tachycardia (VT) in recent years. STAR is a noninvasive and efficient modality, requiring less than 45 minutes per treatment session, capable of targeting arrhythmogenic tissues that are inaccessible through conventional catheter ablation (CA)[1-3]. Notably, numerous recent clinical investigations[4-9] have demonstrated that STAR presents a highly promising approach for the effective management of drug-resistant and recurrent VT.

Nevertheless, STAR has inherent limitations. The treatment target (arrhythmogenic substrates[10]) is subject to movement due to the cardiorespiratory motion of the heart, and the radiation beam cannot dynamically track the target volume (TV). Consequently, to ensure that the TV receives the prescribed radiation dose, the target volume is typically expanded by a margin to form the internal target volume (ITV) to compensate for cardiorespiratory motion. The ITV is further expanded by a 3 to 5 mm margin to the planning target volume (PTV) that accounts for patient positioning and radiation delivery uncertainties[11]. As a result, healthy cardiac tissue surrounding the VT target is also exposed to radiation during treatment, thereby increasing the risk of cardiac radiotoxicity[12-16].

To minimize the risk of cardiac radiotoxicity while effectively delivering the prescribed dose to TV, it is crucial to quantify the patient's cardiorespiratory motion and thus inform the ITV margin definition. In recent years, multiple studies[17-20] have investigated cardiorespiratory motion using different datasets and techniques. In 2016, Ipsen et al.[18] developed a real-time MRI-guided 3D tracking method for non-invasive cardiac radiosurgery to treat atrial fibrillation, targeting the pulmonary vein antrum in the left atrium (LA). The method achieved a mean 3D error of 3.2 mm in the phantoms. The cardiorespiratory motion measured on healthy patients was 16.5 ± 8.0 mm in the Superior-Inferior (SI) direction, 5.8 ± 3.5 mm in the Anterior-Posterior (AP) direction, and 3.1 ± 1.1 mm in the Left-Right (LR) direction. In 2021, Prusator MT, et al.[19] studied the VT targets' cardiorespiratory motion by manually measuring the displacement of heart contours of 11 STAR patients. The 3D target motion range was 6.9 ± 2.6 mm. In 2024, Li G, et al.[20] quantified positional and volumetric variations in arrhythmogenic substrates and cardiac substructures across 12 arrhythmia patients using displacement maximum (DMX) and volume metrics. The study found that the respiratory motion caused a mean substrate target DMX of 0.82 cm (range: 0.32–2.05 cm) and cardiac substructure DMX of 0.35–1.89 cm. The common limitations in these studies included the small sample size[18-20], poor validation in the image registration[19], and use of only healthy subjects[18]. Moreover respiratory 4D CTs (r4DCT) usually show blurry and misaligned artifacts caused by cardiac motion during r4DCT acquisition since cardiac and respiratory motions have different frequencies[21], The cardiac motion artifacts in r4DCTs could make the heart's respiratory motion computed from r4DCTs inaccurate.

Deformable image registration (DIR) is widely used in medical image analysis, including cardiorespiratory motion management studies. In 2010, Myronenko A, et al.[22] developed a Coherent Point Drift (CPD) algorithm that was a probabilistic framework for rigid and non-rigid point set registration. CPD treated point set alignments as a probability density estimation problem with one point set representing Gaussian Mixture Model (GMM) centroids and the other being data points drawn from this GMM. Based on the CPD algorithm, Khallaghi S, et al.[23] proposed GMM-FEM, a novel non-rigid pairwise surface registration method for fusing pre-operative MRI with intra-operative transrectal ultrasound (TRUS) images for prostate interventions. It addressed missing data (up to 30%) in TRUS images due to poor contrast by combining probabilistic correspondences using GMM to treat TRUS surfaces as partial observations and enforcing biomechanical regularization using a Finite Element Model (FEM) to constrain deformations based on tissue properties. In his study, GMM-FEM outperformed the other surface DIR methods, such as TPS-RPM[24], CPD, and ICP-FEM[25]. Key advantages of the GMM-FEM algorithm included robustness to segmentation uncertainties, preservation of volumetric deformation physics (e.g., incompressibility), and comparable performance with full or partial data.

The goal of this study is to compute and analyze the respiratory motion of the heart in r4DCTs of STAR patients so that patient-specific ITV margins can be optimized and thus minimize the radiation to healthy heart tissues. The r4DCT images are usually blurry and have misaligned artifacts caused by unmanaged cardiac motion during r4DCT acquisition and reconstruction, as shown in the example in Figure 1A. It is very challenging to compute only the respiratory motion for the heart while removing the superimposed cardiac motion. To accomplish our goal, we developed a novel image-processing method to compute the respiratory motion of the heart in r4DCTs and filter out the effects caused by the random cardiac motion artifacts.

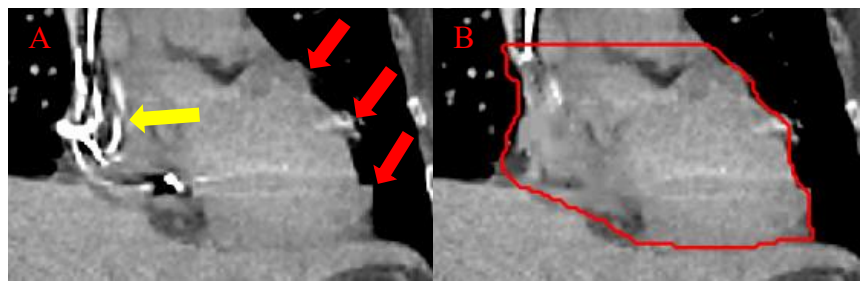

*Figure 1. A respiratory 4D CT slice of a patient. A: Red arrows point to the misalignments caused by cardiac motion. The yellow arrow points to the* Implantable Cardioverter-Defibrillator *lead artifacts. B: The ICD-lead-artifacts-reduced CT slice with the heart contour (red curve) generated by the TheraPanacea ART-plan[26].*

Our image processing method was designed based on multiple important conclusions from previous studies and useful observations:

1) r4DCT images contain both respiratory and cardiac motions of the heart.
2) The respiratory motion is cyclic and continuous in r4DCTs from phase to phase[27].
3) The cardiac motion has a higher temporal frequency than the respiratory motion.
4) Both motions are nearly independent of each other and can be separated, for example, using principal component analysis (PCA)[21,28].

Based on these previous conclusions and observations, we adapted the pairwise GMM-FEM surface-to-surface DIR algorithm[23] into a new groupwise DIR algorithm. We named our new algorithm gCGF, standing for Groupwise CPD-GMM-FEM. CPD stands for Coherent Point Drifting, GMM for Gaussian Mixture Model, and FEM for Finite Element Model. We added two new components to regularize spatial continuity and remove cardiac motion artifacts. The new gCGF algorithm was applied to analyze the respiratory motion of the heart for all participating STAR patients. The main scientific contributions in this work were the new gCGF algorithm and patients' respiratory motion of the heart analysis results. The clinical objective was to deliver quantitative, accurate, and patient-specific respiratory motion data for the heart and VT targets, thereby enhancing motion management in STAR treatments and reducing the risk of cardiac radiotoxicity.

## 2 Methods

### 2.1 Dataset

The r4DCTs of 20 STAR patients previously treated at (institution hidden) before 2022 were analyzed in this study. Patients' DICOM files were retrospectively attained with IRB approval and de-identified prior to analysis. All patients underwent simulation and treatment with 25 Gy delivered in a single fraction under whole-body immobilization and abdominal compression. Each patient's r4DCT contained 10 sets of 3D CT images of the thorax corresponding to 10 phases of an entire respiratory motion cycle. For the cohort of patients, the 3D CT matrix dimensions exhibited a fixed in-plane resolution of 512 x 512 pixels. The number of slices per volume varied between 153 and 279. All scans were acquired at 120 kVp. For r4DCTs, the original slice thickness of all patients was 1.5 mm. All r4DCTs were resampled to an isotropic voxel size of 1.5 mm to allow streamlined data processing. Pixel spacing ranged from 0.9766 mm to 1.2695 mm. Tube current varied from 47 mA to 149 mA, and $CTDI_{vol}$ ranged from 22.4051 mGy to 71.0288 mGy. Further patient CT simulation details regarding the dataset are available in Knutson NC, et al.[11] and Prusator MT, et al.[19].

### 2.2 Segment the heart in 4DCTs

To contour the hearts in r4DCTs, we reduced the metal artifacts in r4DCTs and then segmented the heart using ART-Plan by TheraPanacea[26], an AI-based auto-segmentation software. Figure 1B shows a heart contour example of one phase of an r4DCT. Also shown in Figure 1B, the Implantable Cardioverter-Defibrillator (ICD) lead artifacts were reduced by using an in-house metal-artifact reduction tool (citation hidden) to reduce failures of auto-segmentation using TheraPanacea ART-plan. We manually checked the auto-segmented heart contours and corrected abnormalities. ART-Plan sometimes did not contour the heart correctly. For instance, part of the contour might appear on other organs, such as the lungs and liver, and sometimes the middle part was not contoured. The manual correction ensured that the heart contours were optimally correct and consistent among all r4DCT phases. The shape consistency and manual corrections of heart contours were double-checked by other fellows in our lab. To be explained in later sections, any remaining contouring uncertainties and inconsistencies will be addressed by our gCGF registration algorithm as random artifacts.

### 2.3 gCGF – a groupwise surface-to-surface deformable registration algorithm

While the ICD lead artifacts areas were improved, r4DCTs were still blurry and not ideal for volumetric DIR algorithms, such as pTVreg[29], that operate under the assumption of consistent image appearance and matching voxel intensities. Therefore, we employed a surface-to-surface DIR method in this study to compute the motion of heart contours to investigate the cardiorespiratory motion.

We developed a new surface-to-surface groupwise deformable registration algorithm in this study to iteratively register the heart contours of 10 respiratory phases (phase shapes). We developed gCGF by: 1) converting the pairwise GMM-FEM[22,23] point-cloud deformation registration algorithm into a groupwise setting (note that the pairwise GMM-FEM[22,23] was based on CPD[22]), and; 2) adding a novel principal component filtering (PCF) mechanism and a spatial smoothing mechanism. The pairwise GMM-FEM[22,23] reformulated point-cloud registration as a probability density estimation problem, aiming to maximize the likelihood that one set of points (of phase shapes) is sampled from a probability distribution defined by the other set of points (of the initial shape) and FEM integrates biomechanical regularization by unifying correspondence search and force calculation within a single framework, employing a global probabilistic approach on the target shape's surface[23].

### 2.3.1 Rationale

Rationales for developing the new gCGF algorithm and using it in this study to register the heart in r4DCTs are as follows:

1) The blood pools inside the heart lack trackable information, have inconsistent image intensity due to random cardiac motion artifacts in r4DCTs and ICD wire artifacts, even after metal-artifact reduction. With STAR targets mainly being on the myocardium, registering the heart surfaces is a significantly better choice for studying the motion of the heart and STAR targets than registering the heart volumes. The CPD and GMM components in the gCGF algorithm were good choices for surface registration between two surfaces presented with unpaired surface vertex points.

2) Heart surface delineations, i.e., heart segmentations, in r4DCTs, are affected by random cardiac motion, as demonstrated in Figure 1. A groupwise registration setting is required to allow mechanisms to remove cardiac motion artifacts that are random and uncorrelated with the groupwise cyclic respiratory motion. Standard pairwise registrations are inadequate because they cannot identify and remove motion artifacts uncorrelated with the cyclic motion.

3) Manual corrections on the AI auto-segmented heart contours were necessary for some cases. Uncertainties with auto-segmentations and manual corrections are random and uncorrelated with the regular respiratory cyclic motion. Such uncertainties would be automatically removed by the new gCGF algorithm in the same way as the random cardiac motion artifacts.

4) The 5-dimensional cardiorespiratory MRI data (citation hidden) in our lab showed that the respiratory motions of different parts of the heart were consistently linear, except for the slight compression at inhalation at the bottom of the heart caused by the diaphragm motion. The ideal heart respiratory motion should be semi-rigid and without substantial shape deformation. The FEM component[23] in gCGF would be important to allow the heart to be modeled as a slightly compressible object and ensure the general rigidity.

### 2.3.2 Workflow of the pairwise GMM-FEM

We developed the new gCGF algorithm by adapting the previous algorithm, GMM-FEM[22,23]. The original pairwise GMM-FEM algorithm has the following steps:

1. Perform CPD[22] to calculate the probability distribution matrix $P$ between the phase shape and the initial shape using Equations (1) and (2)[23].

$$\sigma^2 = \frac{1}{3NM} \sum_{n=1}^{N} \sum_{m=1}^{M} \| x_n - y_m \|^2 \qquad (1)$$

$$P(y_m|x_n) = \frac{exp\left(-\frac{\| x_n - y_m \|^2}{2\sigma^2}\right)}{\sum_{j=1}^{M} exp\left(-\frac{\| x_n - y_m \|^2}{2\sigma^2}\right) + c} \qquad (2)$$

where $M, N$ are the number of vertices of the initial and phase shapes, $x_n$ is the n-th vertex of the phase shape, $y_m$ is the m-th GMM centroid of the initial shape, $c$ is the contribution of the outliers and was set to zero in this study. The probability distribution matrix $P$ is an $M$-by-$N$ matrix, and each element is $P_{m,n} = P(y_m|x_n)$.

2. Update FEM node displacements (deformation vector field (DVF)) by minimizing the objective function shown by Equation (3)[23] using FEM parameters, $P$ and $\sigma^2$. Update $\sigma^2$ using Equation (4)[23].

$$Obj(DVF, \sigma^2) = \frac{1}{2\sigma^2} \sum_{m,n=1}^{M,N} P(y_m + DVF_m|x_n) \parallel x_n - y_m \parallel^2$$
$$+ \frac{3 \sum_{m,n=1}^{M,N} P(y_m + DVF_m|x_n)}{2} log(\sigma^2) + \frac{\beta}{2} \cdot DVF^T \circ K \circ DVF \quad (3)$$

$$\sigma^2 = \frac{\sum_{m,n=1}^{M,N} \parallel x_n - (y_m + DVF_m) \parallel^2}{3 \sum_{m,n=1}^{M,N} P(y_m + DVF_m|x_n)} \quad (4)$$

where $K$ is a large sparse matrix dependent on the FEM, a linear material model, $E$ and $v$[23]. "$\circ$" is the matrix multiplication operator. Tikhonov weight $\beta$ represents the level of the volumetric strain energy of the FEMs[23]. $DVF$ is an $M$-by-3 matrix. Update the deformed shape using the DVF ($y_m = y_m + DVF_m$).

3. Replace the initial shape with the updated deformed shape and repeat Steps 1 and 2 until meeting the preset stopping condition[23].

### 2.3.3 Workflow of the gCGF algorithm

The gCGF algorithm has the following three steps, illustrated in Figure 2. The initial shape was registered to the ten phase shapes. More details are described in §2.3.4. A summary of configurable parameters in gCGF is provided in Table 1.

1. Preprocessing: Generate vertices for the initial and phase shapes. Set parameters for the algorithm. Create a FEM model for the initial shape.
2. Iterative groupwise registration: Iteratively deform the initial shape to each of the 10 phase shapes using the GMM-FEM[22,23] workflow described in §2.3.2. Apply spatial smoothing to the deformed shapes. Apply the PCF to the DVFs after each iteration. Update the deformed shapes using the new DVFs. Repeat this step until the stopping condition is met.
3. If the stopping condition of Step 2 is met, switch to the next level of volumetric strain energy and repeat Step 2. The initial shape will be the deformed shape. If all levels of volumetric strain energy are completed, stop the registration, and the deformed shapes are the final registered shapes.

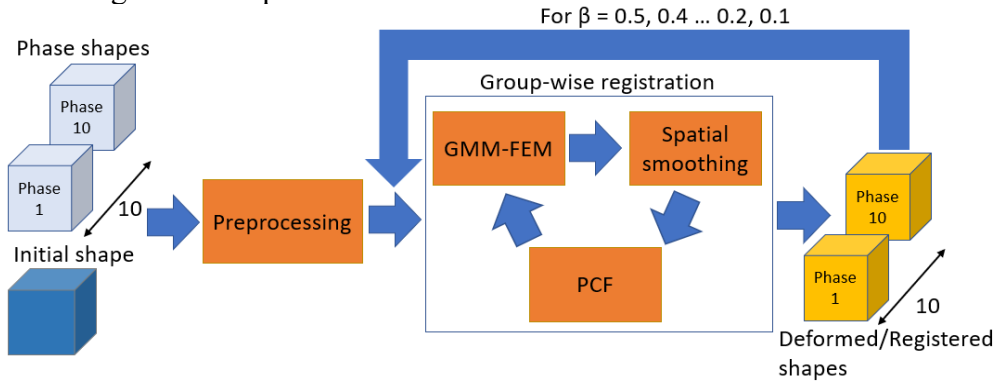

*Figure 2. The workflow of the gCGF algorithm. The ten shapes to be registered were point clouds computed from binary masks of the heart of ten r4DCT phases.*

### 2.3.4 Technique and implementation details

A. Preprocessing: Compute the initial shape by averaging the 10 respiratory heart masks (noted as phase shapes in Figure 2). Transfer the initial and phase shapes into point clouds (vertices). All operations on the "shapes" in the following steps are performed on the vertices of the "shape". Set (initial) values of the parameters. Create a 3D mesh using TetGen[30] and assign physical properties to the 3D mesh to generate a FEM model for the initial shape. The vertices of the initial shape were taken as the FEM nodes and the GMM centroids.

B. Iterative groupwise registration:
   a) For each of the 10 phases, if the percent change (compared to the previous iteration) of $\sigma^2$ of any phase is bigger than the error tolerance and the number of iterations of any phase is no bigger than the maximum number of iterations:
      1. Register the initial shape(s) to the phase shapes using Steps 1 and 2 of §2.3.2. Acquire DVF and update the deformed shapes (GMM centroids) using the DVF.
      2. Apply spatial smoothing: Replace each GMM centroid in the updated deformed shape with the mean average of its neighbors, including itself twice, to acquire a smoothed deformed shape ($y_m^{smooth}$). Update the DVF: $DVF_m = y_m^{smooth} - y_m$.
   b) Reduce cardiac motion artifacts in the ten DVFs from Step a) by applying a PCF step in the phase dimension to remove the motion components in DVFs caused by random cardiac motion[31] in the r4DCTs. This was achieved by applying PCA to the DVFs in the phase dimension and reconstructing the DVFs with the first and second principal components (PCs, the first and second PCs usually correspond to respiratory motion in our observation) and their weights. Update the deform shapes (GMM centroids) using the new DVFs ($y_m^{updated} = y_m + DVF_m^{new}$).
   c) Perform steps a) and b) until the condition mentioned in step a is not met. The updated deformed shapes will be the initial shapes of the next iteration with a new $\beta$ value. Note that each of the updated deformed shapes would be deformed to the corresponding phase in the phase shapes while performing steps a) and b).

C. Perform Step B for each $\beta$ value. If all $\beta$ values have been covered, the deformed shapes are the final registered shapes.

*Table 1. Configurable gCGF parameters.*

| Parameter | Value | Explanation |
| --- | --- | --- |
| Tikhonov weights ($\beta$) | 0.6, 0.5, 0.4, 0.3, 0.2, 0.1 | Level of volumetric strain energy of the FEMs. |
| Number of iterations | 25 | Maximum number of iterations of the groupwise registration. |
| Error tolerance | $10^{-3}$ | See Step B.a). |
| Young's modulus | 5 KPa | Account for FEM's stiffness. |
| Poisson's ratio | 0.49 | Measures the tendency of FEMs to expand sideways when compressed, or contract sideways when stretched. |

## 2.4 Validation of gCGF

Since the ground truth of the heart respiratory motion is unavailable for real patients' cases,

the 4D extended cardiac-torso (XCAT)[32] digital phantom was used to generate 4D heart masks with respiratory motion only, that were treated as the ground truth. Moreover, simulated cardiac motion artifacts were added to the ground truth to create the phase shapes. The initial shape was chosen as the heart mask of the 6$^{th}$ phase (at the end of exhalation) of the same 4D phantom dataset before adding simulated cardiac motion artifacts.

Eight 4D phantom datasets with different levels of simulated cardiac motion artifacts were prepared. The eight datasets were generated with eight combinations of three variables, each with two settings: maximum cardiac artifact magnitude = 2.25 or 4.5 mm, maximum heart bottom expansion factor (the maximum ratio between the radii after and before the expansion) = 1.1 or 1.15, and SI motion range = 10 or 20 mm. Therefore, we had 2×2×2=8 datasets in total. Each dataset consisted of ten 3D phantoms corresponding to ten respiratory phases. More details for generating the datasets are provided in the supplemental document.

gCGF was evaluated under three different settings for each dataset: 1) without PCF and spatial smoothing, 2) with PCF but without spatial smoothing, and 3) with both. For each setting, target registration errors (TRE) were computed based on the following definition:

$$TRE = Mean(\| \overrightarrow{RS} - \overrightarrow{GS} \|) \quad (5)$$

where $\overrightarrow{RS}$ and $\overrightarrow{GS}$ are ten registered and original shapes.

The Second-order difference (SoD) of the registered shapes was computed to evaluate the phase-dimension smoothness of the registration result based on the following equations:

$$SoD = Mean(\|\nabla^2 \overrightarrow{RS}\|) \quad (6)$$

Lower SoD values indicate that the registered shapes have smoother trajectories in the phase dimension. For Equation (6), registered shapes with fewer cardiac motion artifacts would produce a lower SoD value.

## 2.5 Ablation study

We compared the registration accuracy of gCGF with a popular groupwise DIR tool, pTVreg[29]. The 4D phantom was prepared using the procedure described in §2.4 for this task, but only with respiratory motion. To add the misalignment in the r4DCT caused by cardiac motion artifacts, the 3D phantom and heart mask of each phase were divided into ten 3D sections along the SI direction, and each section was shifted in the AP and LR directions by a random distance between ±1.5 mm according to our observation as well as Stevens RRF, et al[17]. The heart masks before the random lateral shifts were used as the ground truth.

For registration with gCGF, the initial shape was the same as that in §2.4 and the phase shapes were point clouds converted from the shifted heart masks. The registration was performed with both PCF and spatial smoothing (the default setting of gCGF). For the registration with pTVreg, the fixed image was the unshifted phantom of phase 6, and the moving images were the shifted phantoms. The 3D smoothing and the phase-dimension smoothing options were enabled in pTVreg. Please refer to the supplemental document for more parameter settings.

The TREs of the gCGF and pTVreg were compared and analyzed. The TREs of the gCGF and pTVreg were computed using Equation (5), where $\overrightarrow{RS}$ was acquired by finding the nearest point in the registered shapes for each point in the ground truth shape for each phase. The registered shape of the pTVreg was acquired by deforming the heart mask of the fixed image by the DVFs and converting the deformed masks into point clouds.

## 2.6 Registration and quantitative analysis of the respiratory motion for STAR patient cases

For each of the 20 STAR patients, the phase shapes and initial shape were heart contours of 10 respiratory phases in r4DCT and the heart contour of the average phase in r4DCT. All registration was conducted with gCGF. After the r4DCT registration, r4DCT DVFs of the heart surface were computed by subtracting the registered shapes from the average registered shape. Since the STAR target contour was only defined on the planning CT for each patient, the heart mask shape of the planning CT was registered to the average registered shape. The resulting DVFs were interpolated to the STAR target surface of the planning CT, and the STAR target shape of the planning CT was deformed to create the STAR target shape of the average registered shape by the interpolated DVFs. Since the ideal heart respiratory motion should be almost rigid, the r4DCT DVFs of the STAR target were acquired by interpolating r4DCT DVFs of the heart surface to the STAR target surface of the average registered shape. For the interpolation, each column of the DVF of each phase was interpolated to the STAR target individually. The sample points were the points of the heart surface, and the query points were the points of the STAR target surface.

The spatial distributions of the respiratory motion magnitude on the surfaces of the hearts and STAR targets of all patients were analyzed. Respiratory motion ranges of individual voxels and the volumetric centroid of the STAR target were calculated.

## 3 Results

### 3.1 Validation, ablation study, and patient registration results

Table 2 shows the TRE and SoD values between the registered shapes and the ground truth shapes for the eight 4D XCAT phantom datasets. The results show that gCGF with PCF and spatial smoothing options produced the lowest TRE and SoD values.

Figure 3 shows the heart center motion trajectories and SoD distributions of phase and registered shapes for a randomly selected patient. Figure 3A shows the heart center trajectory of registered shapes of gCGF with both PCF and spatial smoothing options is almost one-dimensional and less sinuous than other trajectories. Figure 3D shows that registered shapes of gCGF with PCF and spatial smoothing had the lowest SoD values and a more uniform SoD distribution. The observations were true for all patients.

The XCAT phantom and patient registration results in Table 2 and Figure 3 confirmed that gCGF incorporating PCF and spatial smoothing achieved superior accuracy and most effectively enhanced phase-dimension smoothness while reducing cardiac motion artifacts. Table 2 and Figure 3 indicate that PCF, although an outlier rejection method, could improve registration accuracy and phase-dimension smoothness, and account for cardiac motion artifacts in r4DCTs. Adding spatial smoothing could further improve registration accuracy and phase-dimension smoothness. Figure 4 demonstrates that the spatial smoothing option was important for retaining the shape's surface continuity and smoothness after registration.

Table 2. Displacement between moving and initial shapes (in mm), cardiac artifact magnitude (in mm), TRE (in mm), SoD values of the registration results with three conditions averaged across eight datasets.

| Metrics | Value (mm) |
|---|---|
| Mean displacement between the moving and initial shapes before registration | 8.48±6.07 |
| Magnitude of the simulated random cardiac motion artifacts | 0.77±0.40 |
| TREs by gCGF without PCF and spatial smoothing | 1.19±0.88 |
| TREs by gCGF with PCF without spatial smoothing | 0.90±0.67 |
| TREs by gCGF with both PCF and spatial smoothing | 0.63±0.51 |
| Second-order Difference of Ground Truth (without Simulated Cardiac Motion Artifacts) | 1.36 |
| Second-order Difference of Ground Truth with Simulated Cardiac Motion Artifacts | 1.67 |
| Second-order Difference without PCF and spatial smoothing | 1.84 |
| Second-order Difference with PCF without spatial smoothing | 1.33 |
| Second-order Difference with both PCF and spatial smoothing | 1.28 |

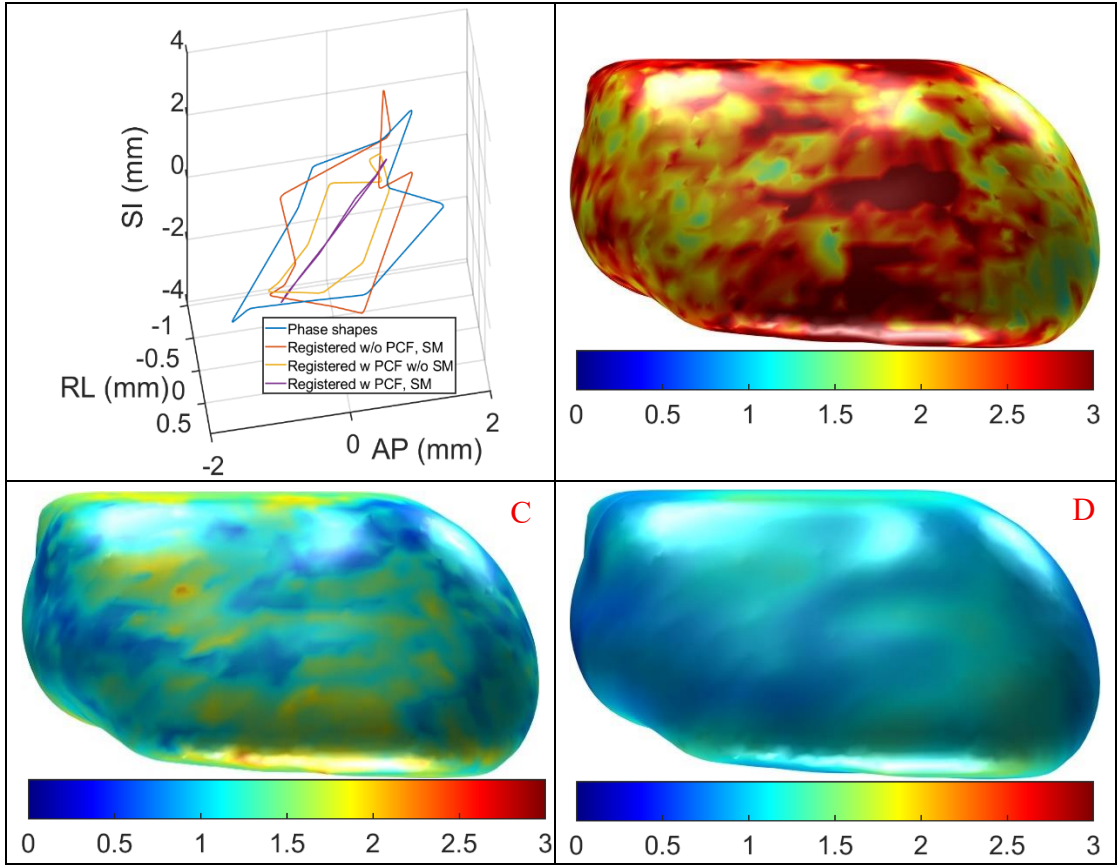

*Figure 3. A) Plots of motion trajectories of heart centers for phase shapes (blue), registered shapes of gCGF without PCF and spatial smoothing (red), registered shapes of gCGF with PCF but without spatial smoothing (yellow), and registered shapes of gCGF with PCF and spatial smoothing (purple). B) Second-order difference (SoD) value distribution on the heart surface of registered shapes of gCGF without PCF or spatial smoothing. C) SoD value distribution on the heart surface of registered shapes of gCGF with PCF but without spatial smoothing. D) SoD value distribution on the heart surface of registered shapes of gCGF with PCF and spatial smoothing. The unit of the color bar is mm.*

| Phase | Initial Shape (static) | Shapes of Respiratory Phases (the Registration Targets) | Registered Shape | |
|---|---|---|---|---|
| | | | Spatial Smoothing Off | Spatial Smoothing On |

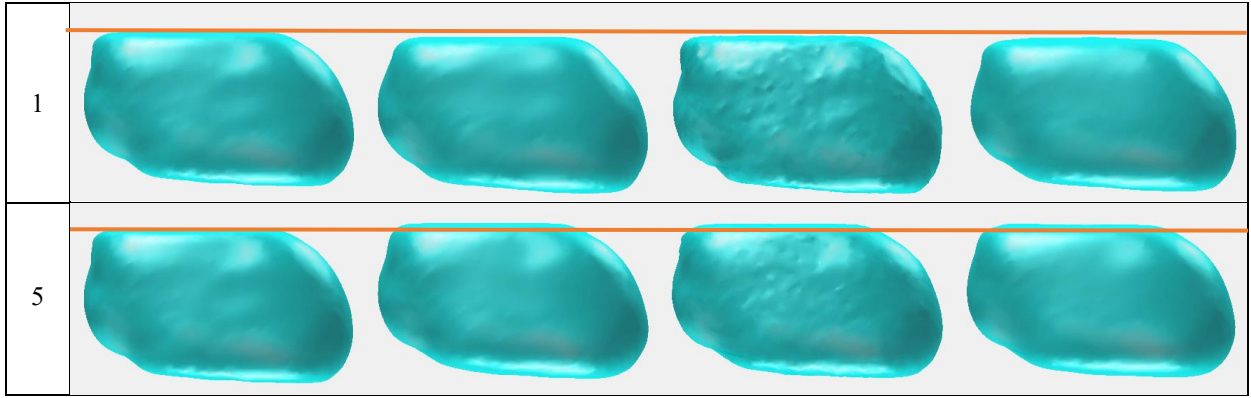

*Figure 4. Examples of unregistered and registered shapes of the two phases of one patient. Registered shapes without spatial smoothing showed rougher surfaces and more variant shapes across phases compared to unregistered shapes and registered shapes with spatial smoothing. The red line represents the position of the top of the initial shape.*

From the comparison study in §2.5, the TREs (averaged over 10 phases) by gCGF and pTVreg were 1.14±0.39 mm and 1.53±1.09 mm, respectively. The average TRE of gCGF was 25% lower than that of pTVreg. A Z-test showed a significant difference between TREs of gCGF and pTVreg across 10 phases (p = 0.0027), indicating gCGF was significantly more accurate than pTVreg in the surface-to-surface DIR task. Please note that these TRE values presented for this comparison study were the surface-to-surface distances instead of point to corresponding point distances. These surface-to-surface TRE values are more generalizable to patient cases for which the ground truth point-to-point distances are unavailable. SoD values were unavailable for the comparison study because point-to-point values were unavailable.

## 3.2 Cardiac respiratory motion analysis

Table 3 shows the maximum, 98th-percentile, and mean values of the heart surface and treatment target surface motion magnitudes in mm, target volumes and locations, and peak-to-peak motion ranges for all STAR patient cases evaluated in this study. The "Max of Max", "98% of Max", and "Mean and Std of Max" values were computed by first calculating the maximum values across ten phases for each voxel, then taking the maximum, 98th percentile, average, and standard deviation values in all voxels. The "95% of All" was computed by acquiring the motion magnitudes of all voxels of all phases and taking their 95th percentile. The "Mean, Std of Mean" values were computed by calculating the mean values across ten phases for each voxel and then taking the mean value of all voxels. It is essential to note that the metrics above were the motion relative to the average registered shapes (acquired by averaging the registered ten shapes) instead of the peak-to-peak motion reported in other studies[17]. "Range" represented the maximum range observed among all voxels on the heart/STAR target surface. For each voxel, its range was defined as the 3D peak-to-peak magnitude.

*Table 3. List of the heart surface and treatment target surface motion magnitudes (in mm), and peak-to-peak motion ranges for all patients studied. All motion (except the Range) in the table was relative to the average position of registered shapes.*

| # | Heart (mm) | | | | | | STAR Target (mm) | | | | | | Vol (cm³) | Location in AHA 17-segment model[33] |
|---|---|---|---|---|---|---|---|---|---|---|---|---|---|---|
| | Max of Max | 95% of All | 98% of Max | Mean, Std of Max | Mean, Std of Mean | Range | Max of Max | 95% of All | 98% of Max | Mean, Std of Max | Mean, Std of Mean | Range | | |
| 1 | 4.7 | 2.6 | 4.2 | 1.9±1.0 | 1.0±0.5 | 9.0 | 3.6 | 2.5 | 3.5 | 1.7±1.0 | 0.8±0.5 | 6.7 | 43.7 | 8,9,14,15 |
| 2 | 3.6 | 2.2 | 3.1 | 1.9±0.5 | 1.0±0.3 | 6.2 | 3.2 | 2.5 | 3.1 | 2.0±0.5 | 1.1±0.3 | 6.2 | 43.5 | 1,2,3,7,8,9 |

| 3 | 5.0 | 2.9 | 4.4 | 2.4±0.9 | 1.3±0.5 | 9.2 | 5.0 | 3.3 | 5.0 | 3.2±0.6 | 1.7±0.3 | 9.2 | 73.5 | 4,5,10,11,15,16 |
|---|---|---|---|---|---|---|---|---|---|---|---|---|---|---|
| 4 | 6.2 | 4.3 | 5.6 | 4.1±0.6 | 2.6±0.4 | 11.4 | 4.5 | 3.8 | 4.4 | 3.7±0.3 | 2.4±0.2 | 8.4 | 72.9 | 4,5,10,11,12 |
| 5 | 6.3 | 4.5 | 6.0 | 4.1±0.8 | 2.5±0.4 | 11.9 | 5.5 | 5.0 | 5.4 | 4.8±0.5 | 2.9±0.3 | 10.2 | 58.6 | 4,9,10 |
| 6 | 5.2 | 3.5 | 4.4 | 3.2±0.6 | 1.8±0.3 | 8.6 | 4.1 | 3.2 | 4.0 | 3.0±0.5 | 1.7±0.3 | 6.8 | 38.3 | 9,10,15 |
| 7 | 6.9 | 4.6 | 6.4 | 4.3±0.9 | 2.4±0.5 | 12.3 | 6.7 | 4.6 | 5.8 | 4.4±0.8 | 2.4±0.4 | 11.8 | 15.2 | 2,3,8 |
| 8 | 4.3 | 2.5 | 3.5 | 2.1±0.7 | 1.2±0.4 | 7.3 | 3.1 | 2.3 | 3.0 | 1.9±0.7 | 1.1±0.3 | 5.3 | 44.9 | 1,2,3,8,9 |
| 9 | 6.9 | 4.4 | 6.0 | 4.4±0.8 | 1.9±0.3 | 11.4 | 5.1 | 3.5 | 5.0 | 3.7±0.7 | 1.6±0.3 | 8.3 | 15.1 | 1,2,6 |
| 10 | 6.0 | 3.7 | 5.3 | 3.5±0.7 | 2.0±0.4 | 10.5 | 5.4 | 3.8 | 5.2 | 3.6±0.8 | 2.0±0.5 | 9.2 | 11.8 | 2,3 |
| 11 | 5.4 | 3.4 | 4.8 | 3.0±0.8 | 1.6±0.4 | 9.9 | 4.8 | 4.1 | 4.8 | 4.0±0.5 | 2.2±0.3 | 8.9 | 11.7 | 5,6 |
| 12 | 4.7 | 2.7 | 3.9 | 2.2±0.8 | 1.0±0.3 | 9.0 | 4.1 | 2.5 | 3.6 | 2.4±0.5 | 1.1±0.2 | 8.0 | 16.3 | 2,3 |
| 13 | 4.9 | 3.2 | 4.0 | 2.7±0.6 | 1.8±0.4 | 9.6 | 3.5 | 3.3 | 3.5 | 2.8±0.5 | 1.9±0.4 | 6.8 | 5.0 | 2 |
| 14 | 6.9 | 3.3 | 5.8 | 2.7±1.2 | 1.4±0.6 | 11.7 | 4.5 | 3.2 | 4.2 | 2.8±0.9 | 1.5±0.5 | 7.4 | 17.8 | 1,6,7,12 |
| 15 | 3.8 | 2.1 | 3.1 | 1.9±0.6 | 1.0±0.3 | 6.3 | 3.0 | 1.9 | 3.0 | 1.8±0.5 | 0.9±0.2 | 4.7 | 18.5 | 3,4,9 |
| 16 | 7.9 | 5.0 | 7.3 | 4.3±1.2 | 2.3±0.6 | 14.7 | 4.4 | 3.6 | 4.3 | 3.5±0.4 | 1.9±0.2 | 8.2 | 13.5 | 15,16,17 |
| 17 | 4.8 | 3.1 | 4.4 | 2.6±0.7 | 1.5±0.4 | 9.1 | 3.1 | 2.2 | 2.6 | 2.1±0.2 | 1.2±0.1 | 5.5 | 16.4 | 1,6 |
| 18 | 4.1 | 2.6 | 3.3 | 2.5±0.5 | 1.3±0.2 | 6.5 | 3.2 | 2.8 | 3.2 | 2.8±0.2 | 1.4±0.1 | 5.6 | 5.9 | 1,6 |
| 19 | 5.8 | 3.8 | 5.6 | 3.3±1.4 | 1.7±0.7 | 9.0 | 4.3 | 2.7 | 4.1 | 2.6±0.7 | 1.4±0.4 | 6.7 | 42.0 | 1,2 |
| 20 | 3.9 | 3.1 | 3.8 | 2.8±0.5 | 1.7±0.3 | 7.4 | 3.2 | 2.8 | 3.2 | 2.6±0.3 | 1.6±0.2 | 6.2 | 48.7 | 1,2,3,7,8,9 |
| Mean | 5.4 | 3.4 | 4.7 | 3.0±0.8 | 1.7±0.4 | 9.6 | 4.2 | 3.2 | 4.0 | 3.0±0.6 | 1.6±0.3 | 7.5 | 30.7 | |

Table 3 indicates that the heart's maximum and mean respiratory motion magnitude ranged from 3.6 to 7.9 mm (Std: 1.2 mm) and 1.0 mm to 2.6 mm (Std: 0.4 mm). The peak-to-peak motion range was from 6.2 to 14.7 mm (Std: 2.2 mm). For VT targets, the max and mean motion magnitude ranges were 3.0 to 6.7 mm (Std: 1.0 mm) and 0.8 to 2.9 mm (Std: 0.3 mm), respectively. The peak-to-peak range was from 4.7 to 11.8 mm (Std: 1.8 mm).

Figure 5 shows the normalized motion magnitude of the heart and the STAR target of each phase, averaged over all patients in 3D, RL, AP, SI, 1st and 2nd PCs directions. Each data point ($Y(ph)$) of one curve in Figure 5 was computed by the following equation:

$$Y(ph) = \frac{1}{N} \sum_{pa=1:N} \frac{M_{ph,pa}}{Ma_{pa}} \quad (7)$$

where $N$ = the number of all patients, $M_{ph,pa}$ is the maximum or mean motion magnitude (relative to the average registered shape) of the heart or STAR target of phase $ph$ of patient $pa$ in the corresponding direction, $Ma_{pa}$ is the maximum motion magnitude (relative to the average registered shape) of the heart of patient $pa$ in the 3D direction.

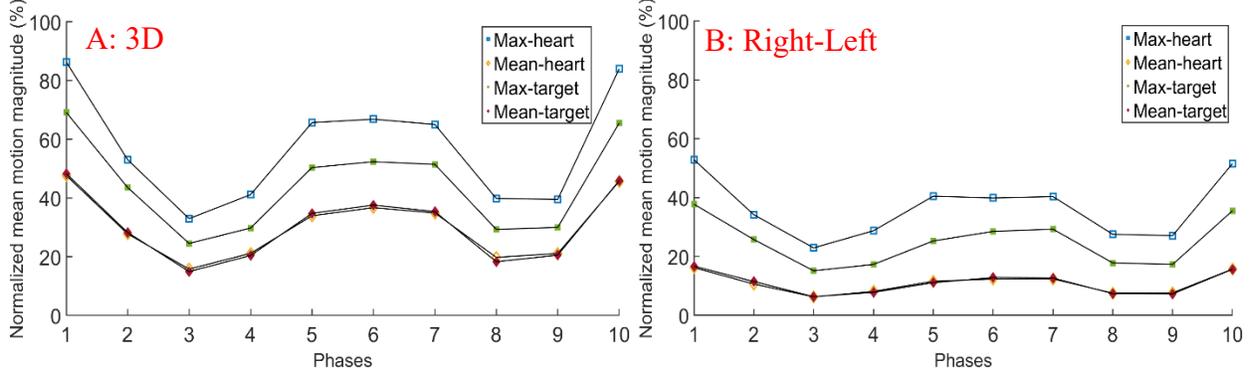

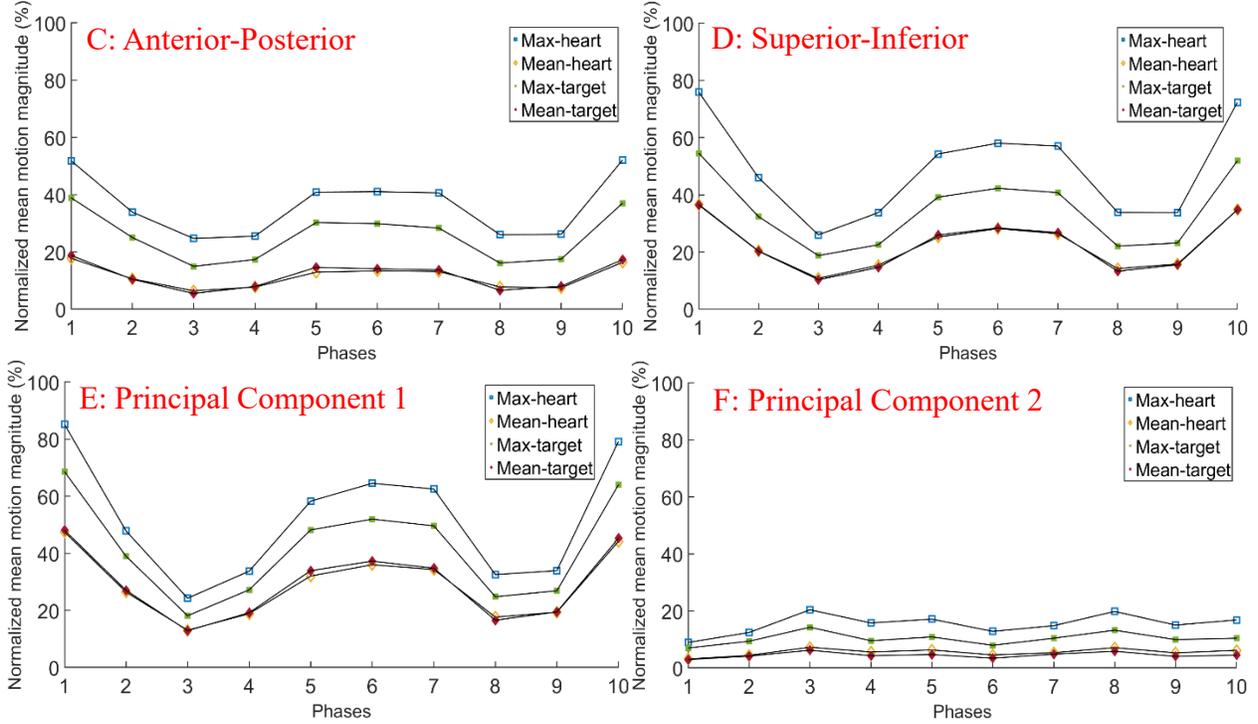

*Figure 5. Normalized mean motion magnitude of the heart and the STAR target of each phase of all patients. In the legend, "Max-heart" means that $M_{ph,pa}$ in Equation (7) is the maximum motion magnitude (relative to the average registered shape) of the heart of phase ph of patient pa when computing the data points on the corresponding curve. The same role applies to other metrics in the legend.*

The six plots in Figure 5 showed that 1) The maximum motion appeared at phases 1 and 10 (end-inhalation). 2) The motion acceleration (changes of motion magnitude) was the largest near phases 1 and 10 (end-inhalation) and the least near phase 6 (end-exhalation). 3) The SI motion was dominant (For Z-tests in heart motion, $p = 0.0484$ between SI and RL directions and $p = 0.0453$ between SI and AP directions. For Z-tests in STAR target motion, $p = 0.0412$ between SI and RL directions and $p = 0.0615$ between SI and AP directions) compared to RL and AP directions, whose motion was comparable ($p = 0.9517$ and $0.8787$ for heart and STAR target motion in Z-tests). 4) The 1st principal motion component was clearly dominant compared to the 2nd component ($p = 0$ for both heart and STAR target in Z-tests), suggesting that the respiratory motion was mainly single-dimensional, along the direction of the 1st principal motion component.

The curves in Figure 5 showed a similar trend, and the normalized mean motion magnitude (values of the y-axis) reached its lowest around phase 3 except for the 2nd principal motion component, indicating that the average position of r4DCT was closest to phase 3.

Figure 6 shows the motion magnitude on the 3D surfaces of the heart and targets for three selected patients. The color assigned to the surfaces was the maximum motion magnitude (relative to the average registered shape) of the ten respiratory phases. The color bars were set to have a maximum value of 6 mm. Patients 2, 11, and 16 were selected to represent patients with "Max of Max" lower than 4 mm, between 4 mm and 6 mm, and larger than 6 mm. Note that the heart shapes shown in the top row of Figure 6 were the average registered shapes.

|  | Patient 2 | Patient 11 | Patient 16 |
|---|---|---|---|

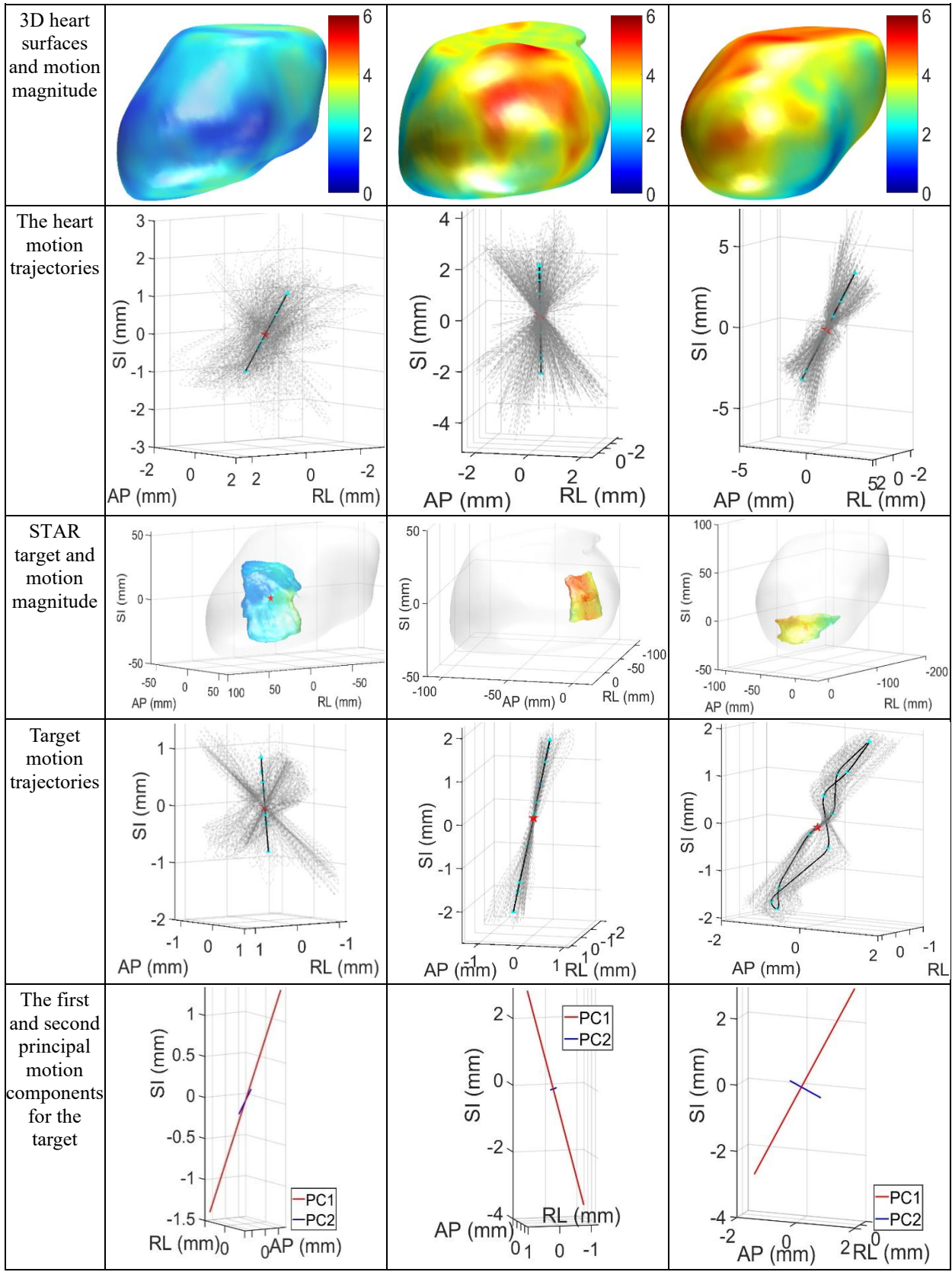

*Figure 6. The motion of the heart and STAR targets in the respective average position images for the three*

*selected patients. In the 2$^{nd}$ and 4$^{th}$ rows, the black curves are the 3D trajectories of the centroid positions of the heart/target of ten phases. Green triangles represent the heart/target centroid positions of each phase. The red pentagram represents the centroid of the STAR target's coordinates in the average registered shape. Each dashed curve represents a voxel's trajectory on the heart/target surface. The dashed curves were put together by translating them by the distance between the pentagram and the corresponding voxel of the average registered shape. Pictures in the 5$^{th}$ row were created by applying PCA to the target DVFs, reconstructing DVFs corresponding to the 1$^{st}$ and 2$^{nd}$ PCs, and plotting trajectories for the 1$^{st}$ and 2$^{nd}$ PCs using the reconstructed DVFs.*

The motion of Patient 2 was small and the motion trajectories of the voxels pointed to different directions due to the small and generally uncoordinated motion. Patients 11 and 16 had larger motion. The heart centroid usually moved in an almost straight line. Moreover, we found that some parts of the heart had larger motion than others, which indicated the heart's respiratory motion included rotation or tilting. The 2$^{nd}$ and 4$^{th}$ rows of Figure 6 shows that different voxels in the heart and target moved differently. The 5$^{th}$ row confirmed that the 1$^{st}$ PC of the target motion is dominant, which is true for all patients except Patient 17, whose 2$^{nd}$ PCs' motion was also obvious (the range of PC2 is 69.6% of PC1), but the 1$^{st}$ PC's target motion range is still larger.

## 4 Discussion

In this study, we developed gCGF to register the heart contours of ten phases of r4DCT and to compute the respiratory motion of the heart. The heart contours used for registration were acquired by an AI-based auto-segmentation followed by manual corrections. The contouring uncertainties were treated as cardiac motion artifacts and accounted for during gCGF registration. For any new patient, the AI cardiac segmentation results will likely be used without secondary manual validation. A future study might be needed to test whether systematic AI segmentation errors on the low-quality r4DCTs could still impact the 3D respiratory motion analysis, even after random cardiac motion artifacts are accounted for by the gCGF algorithm. We did not perform the registration with "gCGF without PCF, with spatial smooth" in the validation study (though the results would look more thorough with this option), since spatial smoothing alone apparently could not handle cardiac motion artifacts.

The observations from Figure 5 and Figure 6 are pertinent to clinical practice. Figure 5 showed that SI motion is the primary challenge for heart motion management, exhibiting a larger motion magnitude than RL/AP. However, some patients also had large AP motion, such as Patient 16 of Figure 6, whose 5$^{th}$ row indicated that the target respiratory motion is mainly 1D. Therefore, the 1$^{st}$ PC's (sometimes 1$^{st}$ and 2$^{nd}$ PCs') direction should be focused when defining the target respiratory motion margin.

The mean values of the maximum target motion ranges of all patients in RL, AP, and SI directions in this study were 4.3 mm, 4.4 mm, and 6.4 mm, which were larger than those of the previous study, Prusator MT, et al, 2021[19](3.9 mm, 4.1 mm, and 4.7 mm). Note that the ranges in Prusator MT, et al were acquired by manually measuring the displacement between the heart contours in different phases.

The limitations of this study include: 1) The registration results only contained motion on the heart surface since gCGF is a surface-to-surface DIR algorithm. 2) The DVFs of the STAR targets were acquired by interpolating the DVFs of the heart surface to the target surface since target contours of r4DCTs were unavailable.

The future work includes optimizing the PCF method to make it reduce cardiac motion artifacts and improve phase-dimension smoothness when the respiratory motion is not dominant in the

r4DCT, exploring the possibility of registering tumors in the proximity of the heart with gCGF, and designing more realistic ground truth shapes for validation.

## Conclusion

The proposed gCGF algorithm could register r4DCTs accurately and account for cardiac motion artifacts while improving phase-dimension smoothness. Although the SI motion was confirmed to be more dominant compared to motion in RL and AP directions, STAR planning should prioritize the 1$^{st}$ PC's respiratory motion for more accurate and patient-specific target margins.